\documentclass[traditabstract,referee]{aa} % for a referee version
\usepackage{natbib,graphicx,hyperref} %,amsmath,amssymb,graphicx,hyperref}
\usepackage[varg]{txfonts}
\usepackage{xcolor}
\newcommand{\rsun}{R_{\odot}}
\newcommand{\RB}{R_{\rm b}}
\begin{document} 

   \title{An update of Leighton's solar dynamo model}

   \author{R.~H.~Cameron
          \inst{1}
          \and
          M.~Sch{\"u}ssler\inst{1}
          }

   \institute{Max-Planck-Institut fÃ{\"u}r Sonnensystemforschung
     Justus-von-Liebig-Weg 3, 37077 G{\"o}ttingen\\
     \email{cameron@mps.mpg.de}
             }

   \date{Received ; accepted }

% \abstract{}{}{}{}{} 
% 5 {} token are mandatory
 
\abstract 
  {In 1969, Leighton developed a quasi-1D mathematical model
  of the solar dynamo, building upon the phenomenological scenario of
  Babcock published in 1961. Here we present a modification and
  extension of Leighton's model. Using the axisymmetric component
  (longitudinal average) of the magnetic field, we consider the radial
  field component at the solar surface and the radially integrated
  toroidal magnetic flux in the convection zone, both as functions of
  latitude. No assumptions are made with regard to the radial location
  of the toroidal flux. The model includes the effects of (i) turbulent
  diffusion at the surface and in the convection zone, (ii) poleward
  meridional flow at the surface and an equatorward return flow
  affecting the toroidal flux, (iii) latitudinal differential rotation
  and the near-surface layer of radial rotational shear, (iv) downward
  convective pumping of magnetic flux in the shear layer, and (v) flux
  emergence in the form of tilted bipolar magnetic regions treated as a
  source term for the radial surface field. While the parameters
  relevant for the transport of the surface field are taken from
  observations, the model condenses the unknown properties of magnetic
  field and flow in the convection zone into a few free parameters
  (turbulent diffusivity, effective return flow, amplitude of the source
  term, and a parameter describing the effective radial shear).
  Comparison with the results of two-dimensional flux transport dynamo
  codes shows that the model captures the essential features of these
  simulations. We make use of the computational efficiency of the model
  to carry out an extended parameter study. We cover anextended domain of
  the four-dimensional parameter space and identify the parameter ranges
  that provide solar-like solutions. Dipole parity is always preferred
  and solutions with periods around 22 years and a correct phase
  difference between flux emergence in low latitudes and the strength of
  the polar fields are found for a return flow speed around
  2~m$\cdot$s$^{-1}$, turbulent diffusivity below about
  80~km$^2\cdot$s$^{-1}$, and dynamo excitation not too far above the
  threshold (linear growth rate less than 0.1~yr$^{-1}$).}  


%   \titlerunning{Babcock-Leighton solar dynamo model}
   \maketitle
%
%-------------------------------------------------------------------

\section{Introduction}

\citet{Babcock:1961} proposed a scenario describing the cyclic
solar dynamo in terms of a consistent physical approach based on
observational results. These were basically the (latitudinal)
differential rotation, the polarity rules of sunspot groups and their
systematic tilt with respect to the East-West direction, the reversals
of the global dipole field in the course of the solar cycle together
with the relationship between its orientation and the sunspot
polarities. In Babcock's scenario, the poloidal magnetic field
represented by the global dipole is wound up by differential
rotation. Loops of the resulting azimuthal field rise owing to magnetic
buoyancy and break through the surface, thus forming sunspot groups and
bipolar magnetic regions (BMRs). These are tilted in the observed sense
because of the poloidal field component provided by the global dipole.%
\footnote{Interestingly, Babcock does not comment on the fact that the
  latitude dependence of the resulting BMR tilt according to his model
  is in conflict with observation (Joy's law). Nevertheless, although he
  neglected the effect of the Coriolis force on rising flux bundles, his
  original scenario can work as a proper dynamo owing to the
  non-axisymmetric character of flux emergence.}
The BMR tilt leads to a preferential cancellation of leading-polarity
magnetic flux across the equator, leading to an amount of net
following-polarity flux on each hemisphere. As this flux spreads over
the hemisphere and migrates polewards, it eventually cancels and
reverses the global dipole field. This field is the source of the
(reversed) toroidal field of the next activity cycle, thus leading to a
22-year magnetic cycle.

\citet[][hereafter referred to as L69]{Leighton:1969} condensed
Babcock's scenario into a mathematical model in the form of two coupled
partial differential equations (in time and latitude), representing the
toroidal and poloidal components of the azimuthally averaged
(axisymmetric) magnetic field. He added radial shear to the differential
rotation and treated the transport of surface magnetic flux by
supergranular flows in terms of a diffusion model \citep[random walk of flux
elements,][]{Leighton:1964}.

The dynamo model of \citet{Parker:1955} differs conceptually from that
of Babcock and Leighton in the mechanism envisaged for the reversal and
regeneration of the poloidal field. He considers correlations between
small-scale convective motions brought about by the systematic effect of
the Coriolis force in a rotating system. This approach was later
systematically worked out by the Potsdam group in terms of mean-field
turbulence theory \citep[see, e.g.,][]{Krause:Raedler:1980}, leading to
the $\alpha$-effect paradigm (`turbulent dynamo'). Although the
resulting dynamo equations (in the 1D case with only latitudinal
dependence) are mathematically similar to the corresponding equations of
Leighton \citep[see][]{Stix:1974}, this must not obscure the fundamental
conceptual difference between the two approaches: while the turbulent
dynamo relies on the collective effect of small-scale correlations
throughout the convection zone, the Babcock-Leighton (BL) approach is
based upon the actually observed properties and evolution of active
regions at the solar surface. In particular, it is the big bipolar
regions emerging not too far from the equator that contribute most to
the (re)generation of the poloidal field in the BL model.

The structure of the dynamo equations for both approaches permits cyclic
solutions with propagating dynamo waves resembling the latitudinal
propagation of the activity belts in the course of the solar cycle,
provided that rotational shear is mainly in the radial direction.%
\footnote{Actually, in \citet{Leighton:1969} there is a model with a
  latitudinally propagating dynamo wave in the absence of radial
  shear. However, this result is due to an unphysical feature in his
  formalism, which leads to a violation of the condition
  $\nabla\cdot{\bf{B}}=0$ (see Appendix).}
When helioseismic observations proved this assumption wrong
\citep{Duvall:etal:1984} and a systematic poleward meridional flow was
detected at the surface \citep[starting with][]{Howard:1979}, the
propagation of the activity belts was alternatively ascribed to
equatorward magnetic flux transport by a deep latitudinal return flow
towards the equator \citep{Wang:etal:1991, Choudhuri:etal:1995}.  In
parallel, numerical models treating the evolution of the large-scale
surface magnetic field as a result of the combined effects of flux
emergence in tilted bipolar regions, differential rotation, meridional
flow, and supergranular random walk (treated as turbulent diffusion)
successfully reproduced observations, including the reversals of the
polar fields during cycle maxima, and thus validated a central concept
of the Babcock-Leighton scenario \citep[for reviews,
see][]{Mackay:Yeates:2012,Jiang:etal:2014,Wang:2016}.  This led to a
revival of interest in this concept \citep{Wang:Sheeley:1991} with a
first update of the L69 dynamo model by \citet{Wang:etal:1991}.
Eventually, this brought about the development of spatially
two-dimensional Babcock-Leighton-type flux-transport dynamo (FTD) models
\citep[see reviews by][]{Charbonneau:2010, Karak:etal:2014}, which
recently have been extended to include 2D surface transport
\citep{Lemerle:etal:2015} or even to spatially three-dimensionsal
treatment \citep{Miesch:Teweldebirhan:2015}.  Such models are able to
reproduce key features of the solar activity cycle if their parameters
are properly `tuned'. In particular, the results depend strongly on the
assumptions about the turbulent diffusivity and about the spatial
structure of the meridional circulation in the convection zone
\citep{Charbonneau:2010, Karak:etal:2014}.

Recently, additional information concerning the operation of the solar
dynamo has become available. \cite{Wang09} and \cite{Munoz-Jaramillo13}
    showed that the polar fields (or proxies thereof, such as the minima of the open heliospheric flux or of geomagnetic activity) at activity minimum is
a good proxy for the strength of the following solar cycle. Thereafter
\citet{Cameron:Schuessler:2015} showed by a
mathematical argument that the net toroidal flux in a solar hemisphere that
produced by differential rotation is determined by the emerged
magnetic flux at the solar surface. They also found that the latitudinal
differential rotation is the by far dominant generator of net toroidal
flux, while the near-surface shear layer
\citep{Thompson:etal:1996,Barekat:etal:2014} plays only a minor
role. Considering the observed distribution of magnetic flux at the
solar surface, these authors also showed that the poloidal flux relevant
for the generation of net toroidal flux is mostly connected to polar
fields, i.e., the axial dipole field. These results strongly support the
validity of the Babcock-Leighton scenario.

\citet{Cameron:Schuessler:2016} used the observed properties of the
sunspot butterfly diagrams observed since 1874 to infer the turbulent
magnetic diffusivity affecting the toroidal field in the convection
zone. They found the turbulent diffusivity to be in the range
150--450~km$^2$s$^{-1}$, which is consistent with simple estimates from
mixing-length models and thus puts the solar dynamo in the
`high-diffusivity' regime \citep[cf.][]{Charbonneau:2010}.

Given the relevant observational results obtained since 1969, namely,
the systematic poleward meridional flow at the surface (which actually
was already conjectured by Babcock in his 1961 paper), the measurement
of the differential rotation in the convection zone by helioseismology,
together with the inference of a high turbulent diffusivity, a new
update of the L69 model seems in order. The simplicity of the spatially
1D approach minimizes the number of free parameters (and functions) and
yet allows us to include all relevant physical ingredients. Since it is
computationally inexpensive, extensive parameter studies can be carried
out and very long runs covering thousands of cycles can be performed,
thus permitting statistical studies, e.g. to compare with long-term
results obtained from cosmogenic isotopes \citep{Usoskin:2013}.

This paper is structured as follows.  In Sec.~\ref{sec:model} we outline
the model assumptions and the mathematical formulation of our model. As
a validation, the results of the model are compared to those of 2D flux
transport dynamo codes in Sec.~\ref{sec:FTD}. Results of an extended
parameter study are shown in Sec.~\ref{sec:parameter_study}. Our
conclusions are given in Sec.~\ref{sec:conclusions}. The Appendix points
out an unphysical feature in Leighton's original formulation that leads
to latitudinal dynamo waves in the absence of radial shear.

\section{Model}
\label{sec:model}

\subsection{The model of Leighton (1969)}
\label{subsec:Leighton}

The basic concept of the original L69 model is to condense the Babcock
scenario into a system of two partial differential equations (depending
on time and latitude) for the azimuthally averaged radial magnetic field
at the surface, $B_r$, and azimuthal field, $B_\phi$, located in a
narrow layer below the surface.  While the surface field undergoes a
random walk described as turbulent diffusion, there is no diffusion for
$B_\phi$, which is built up by latitudinal differential rotation as
observed at the surface and latitude-dependent radial shear in a narrow
sub-surface layer. The source of the radial field is provided by flux
eruption in tilted bipolar magnetic regions, which is described by a
double-ring formalism. Flux eruption is assumed to require a minimal
azimuthal field strength and contains a random element.

\subsection{The updated model}
\label{subsec:updated}

The model described here follows the general concept of the L69
model. Modifications and new features reflect results obtained
after the publication of the original model. The key features of our
updated model are summarized in the following list.
\begin{itemize}
\item{As second variable besides the radial field at the surface we
  consider the radially integrated toroidal flux (per unit length in the
  latitudinal direction) in the convection zone.}
\item{Radial surface field and toroidal flux in the convection zone are
  both affected by turbulent diffusion and meridional flow. The
  turbulent diffusivities at the surface and in the convection zone can
  take different values. There is no diffusive emergence of toroidal
  flux at the surface and no radial transport of toroidal flux at the
  bottom of the convection zone. The meridional flow is poleward at the
  surface and equatorward in the part of the convection zone where the
  toroidal flux resides.}
\item{The rotation rate varies latitudinally according to the surface
  rate and radially according to the near-surface shear layer (NSSL)
  found by helioseismology \citep{Thompson:etal:1996}.}
\item{Effects of downward convective pumping are included.} 
\item{Flux emergence in tilted BMRs is described by a source term for
  the radial surface field formally similar to the $\alpha$-term of
  mean-field theory. Its specific form reflects the latitude dependences
  of the Coriolis force and of the azimuthal length of toroidal flux
  bundles. There is no threshold toroidal field strength for flux
  emergence.}
\item{Flux emergence does not deplete the reservoir of toroidal
  subsurface flux \citep{Parker:1984}. Toroidal field is removed by
  cancellation and `unwinding' \citep{Wang:Sheeley:1991,
  Cameron:Schuessler:2016}.}
\end{itemize}
In contrast to the earlier update to the L69 model by
\citet{Wang:etal:1991}, who already incorporated meridional flow and
diffusion of the toroidal field, our treatment does not require us to
make assumptions about the location and structure of the toroidal flux
in the convection zone. Furthermore, we now include the effect of the
NSSL as well as of downward convective pumping.

\subsection{Equations}
\label{subsec:equations}

The mathematical model consists of two coupled equations representing
the evolution of the axisymmetric component (azimuthal average) of the
magnetic field a a function of time and colatitude under the influence
of differential rotation, meridional flow, and turbulent diffusion.  The
first equation characterizes the evolution of the radially integrated
toroidal flux in the convection zone per unit colatitude. The second
equation describes the evolution of the radial component of the field at
the visible solar surface, $B_{r,{R_\odot}}$, in terms of the surface
flux transport (SFT) model.

In spherical polar coordinates $(r,\theta,\phi)$, the $\phi$-component of
the induction equation for the mean (azimuthally averaged) magnetic
field is given by
\begin{eqnarray}
  \frac{\partial B_{\phi}}{\partial t}&=&\frac{1}{r}
  \left\{\frac{\partial \left[r \left(U_{\phi} B_r
        -U_r B_{\phi}\right)\right] }{\partial r}
        +\frac{\partial \left(U_{\phi} B_{\theta} 
        - U_{\theta} B_{\phi}\right)}{\partial\theta}\right. \nonumber \\
   & & \left. +\frac{\partial}{\partial r} 
        \left[ \eta\frac{\partial(r B_{\phi})}{\partial r} \right] 
        + \frac{\partial}{\partial\theta} 
        \left[ \frac{\eta}{r \sin \theta}
        \frac{\partial (B_\phi\sin\theta)}{\partial\theta}\right]\right\}\,.
\label{eq:induction}
\end{eqnarray}
The components of the mean velocity correspond to differential rotation,
$U_\phi=\Omega(r,\theta)r\sin\theta$, mean meridional flow, $U_\theta$,
and convective pumping, $U_r$. The symbol $\eta(r)$ represents the turbulent
diffusivity, for which we allow a radial dependence.  We consider the
toroidal flux (per unit colatitude) integrated radially between the
bottom of the convection zone at $r=\RB$ and the solar surface, viz.
\begin{equation}
 b(\theta,t) = \int_{\RB}^{R_{\odot}}B_{\phi}(r,\theta) r \mathrm{d}r\,.
\label{eq:bdef}
\end{equation}
Integrating Eq.~(\ref{eq:induction}) we obtain
\begin{eqnarray}
  \frac{\partial b}{\partial t}
  &=&\int_{\RB}^{R_{\odot}} \sin\theta \frac{\partial(\Omega r^2 B_r)}
       {\partial r} \mathrm{d}r + \frac{\partial}{\partial \theta} 
       \int_{\RB}^{R_{\odot}} \Omega \sin\theta \,
            B_{\theta}\, r \mathrm{d}r \nonumber \\
  & &  -\int_{\RB}^{R_{\odot}} \frac{\partial(r U_r B_{\phi})}
             {\partial r} \mathrm{d}r
       -\int_{\RB}^{R_{\odot}} \frac{\partial (U_{\theta} B_{\phi})}
             {\partial \theta} \mathrm{d}r  \nonumber \\
  & &  +\int_{\RB}^{R_\odot} \frac{\partial}{\partial r} 
             \left[\eta \frac{\partial (r B_{\phi})}
             {\partial r} \right] \mathrm{d}r   \nonumber \\
  & &  +\int_{\RB}^{R_\odot} \frac{\partial}{\partial \theta} 
             \left[ \frac{\eta}{r \sin \theta}
             \frac{\partial (B_\phi \sin\theta)}{\partial\theta}\right] 
             \mathrm{d}r \,.
\label{eq:integrated}
\end{eqnarray}
The first two terms on the right-hand side represent the generation of
toroidal flux by radial and latitudinal differential rotation,
$\Omega(r,\theta)$. The third and fourth terms refer to transport of
toroidal flux by convective pumping (radial) and meridional flow
(radial and latitudinal). The last two terms describe the effect of
(turbulent) diffusion. 

The formulation in terms of integrated toroidal flux avoids the need to
specify a storage region for the toroidal flux. In order to further
evaluate Eq.~(\ref{eq:integrated}) we nevertheless have to make some
assumptions about the magnetic field and flows in the convection
zone. These are:

\begin{itemize}
\item{Magnetic flux does not penetrate into the radiative interior
      beneath the convection zone. This entails $B_r=0$ at $r=\RB$,
      together with the absence of radial transport ($U_r=0$ at
      $r=\RB$), as well as by radial diffusion ($\partial r
      B_\phi/\partial r = 0$ at $r=\RB$). These are reasonable
      assumptions considering the strong entropy barrier at the
      interface to the radiative interior.}
\item{Downward convective pumping expels the horizontal field components in
      the near-surface shear layer (NSSL) located in the uppermost part
      of the convection zone, so that $B_\theta = B_\phi=0$ in the NSSL
      \citep[cf.][]{Karak:Cameron:2016}. This entails 
      $B_\phi = \partial (r B_\phi)/\partial r = 0$ at $r=\rsun$).}
\item{The poloidal field ($B_r, B_\theta$) does not penetrate into the
      convection zone part of the tachocline. This assumption is
      justified since, firstly, most of the tachocline is located below
      the convection zone in lower latitudes and, secondly, unless
      convective pumping is extremely strong, only a minor part of the
      radial magnetic flux actually reaches down to the convection zone
      part of the tachocline in the higher latitudes. Furthermore, it is
      unclear whether the tachocline shear can at all generate a
      substantial amount of toroidal flux \citep{Vasil:Brummell:2009,
        Spruit:2011}, and the presence of a tachocline does not prevent solar-like cyclic
        dynamo action in fully convective stars \citep{2016ApJ...830L..27R, 2016Natur.535..526W}}.
\item{The radial shear vanishes between the NSSL and the top of the
      tachocline. In low and middle latitudes, this is justified by the
      results of helioseismology. Some deviation from this assumption is
      taken into account by introducing an additional parameter,
      $\epsilon$. Anyway, such deviations can only affect the
      latitudinal distribution, but not the total amount of the net
      toroidal flux in each hemisphere \citep{Cameron:Schuessler:2015}.}
\end{itemize}

These assumptions locate the generation of toroidal flux by radial
shear in the NSSL while the generation by latitudinal shear occurs in
the bulk of the convection zone below the NSSL. 
Using these assumptions, the generation of toroidal flux by differential
rotation, which is described by the first two terms on the right-hand
side of Eq.~(\ref{eq:integrated}), has been worked out by
\citet[][Appendix]{Cameron:Schuessler:2016}, who obtained
\begin{eqnarray}
  \left( \frac{\partial b}{\partial t}\right)_{\rm DR}  &=& \sin\theta\,
  R_\odot^2 B_{r,{R_{\odot}}} \left( \Omega_{R_\odot} -
  \Omega_{R_{\rm NSSL}} \right) \nonumber \\
  & & -\frac{\partial \Omega_{R_{\rm NSSL}}}{\partial \theta}
    \int_0^{\theta} \sin\theta\, R_{\odot}^2 B_{r,{R_{\odot}}}
  \mathrm{d}\theta \,,
\end{eqnarray}
where $\Omega_{R_{\rm NSSL}}(\theta)$ and $\Omega_{\rsun}(\theta)$ are
the latitudinal angular velocity profiles at the bottom of the NSSL,
$R_{\rm NSSL}$, and at the surface, respectively. For the surface rotation
rate we use the synodic rate for magnetic fields determined by
\citet{Hathaway:Rightmire:2011},
\begin{eqnarray}
\label{eq:omega_rsun}
\Omega_{R_{\odot}}(\theta) &=& 14.30-1.98 \cos^2\theta  \nonumber \\
& & -2.14 \cos^4\theta \mbox{\hspace{0.2cm}} [ ^{\circ}/\mbox{day} ]\,.
\end{eqnarray}
To define the latitude profile of the angular velocity at the bottom of
the NSSL we start from the near-surface profile determined from
helioseismology by \citet{Schou:etal:1998} and add a
latitude-independent value of $0.53^{\circ}/$day following
\citet{Barekat:etal:2014}, who found that radial shear in the NSSL is
independent of latitude. Thus we obtain
\begin{eqnarray}
\label{eq:omega_rNSSL}
\Omega_{R_{\rm NSSL}}(\theta) &=& 14.18-1.59 \cos^2\theta \nonumber \\
 & &-2.61
  \cos^4\theta + 0.53 \mbox{\hspace{0.2cm}} [ ^{\circ}/\mbox{day} ]\,.
\end{eqnarray}
Next we consider the transport terms on the right-hand side of
Eq.~(\ref{eq:integrated}). The third term, describing radial advection,
can be directly integrated and vanishes since $U_r=0$ at the bottom and
$B_\phi=0$ at the top.  The fourth term (latitudinal advection) is
rewritten in the form
\begin{eqnarray}
-\frac{\partial}{\partial\theta} \int_{\RB}^{R_{\odot}} 
     U_{\theta} B_{\phi} \mathrm{d}r & = &  
-\frac{\partial}{\partial\theta} \left[ \overline{ \left(
      \frac{U_\theta}{r} \right) }  \int_{\RB}^{R_{\odot}} 
  B_{\phi} r \mathrm{d}r \right] \equiv 
- \frac{\partial}{\partial\theta} 
      \left(\frac{ V_\theta}{\rsun} \, b \right) \,,
\label{eq:adv_mf}
\end{eqnarray}
where $V_\theta=R_\odot \overline{U_\theta/r}$ is a weighted average of
the latitudinal meridional flow in the depth range where the toroidal
flux resides. We assume this to represent the equatorward return flow of
the meridional flow at the surface and write, for simplicity, 
$V_\theta = V_0 \sin(2\theta)$, where $V_0>0$ is a free parameter of the
model. 

Finally, we consider the diffusion terms in Eq.~(\ref{eq:integrated}),
i.e., the fifth and sixth terms on its right-hand side. The fifth term
vanishes since we assume that there is no diffusive flux of toroidal 
field over the radial boundaries of the convection zone. To rewrite the
sixth term we assume that the turbulent diffusivity has a radial
profile given by $\eta(r)=\eta_0 (r/\rsun)^2$. This leads to
\begin{eqnarray}
\int_{\RB}^{R_\odot} \frac{\partial}{\partial\theta} & & 
             \left[ \frac{\eta}{r \sin \theta} 
             \frac{\partial (B_\phi \sin\theta)}{\partial\theta}\right] 
             \mathrm{d}r   \nonumber \\ 
 & & =  \frac{\eta_0}{\rsun^2} \frac{\partial}{\partial\theta}\left[
             \frac{1}{\sin\theta} \int_{\RB}^{R_\odot}
             \frac{\partial (\sin\theta B_\phi r)}{\partial\theta}
             \mathrm{d}r\right]   \nonumber \\ 
  & & = \frac{\eta_0}{\rsun^2} \frac{\partial}{\partial\theta}\left[
             \frac{1}{\sin\theta}\frac{\partial}{\partial\theta}
             (\sin\theta\,b)\right] \,.
\label{eq:diffusion}
\end{eqnarray}
Adding all contributions together, we obtain
\begin{eqnarray}
\frac{\partial b}{\partial t}
  &=& \sin\theta\,
  R_\odot^2 B_{r,{R_{\odot}}} \left( \Omega_{R_\odot} -
  \Omega_{R_{\rm NSSL}} \right) \nonumber \\
 & &  -\frac{\partial \Omega_{R_{\rm NSSL}}}{\partial \theta}
    \int_0^{\theta} \sin\theta\, R_{\odot}^2 B_{r,{R_{\odot}}}
  \mathrm{d}\theta   \nonumber \\ 
  & & -\frac{1}{\rsun} \frac{\partial}{\partial\theta} 
      \left( V_\theta \, b \right)
      + \frac{\eta_0}{\rsun^2} \frac{\partial}{\partial\theta}\left[
        \frac{1}{\sin\theta}\frac{\partial}{\partial\theta}
        (\sin\theta\,b)\right] \,.
\label{eq:btor_1}
\end{eqnarray}

To describe the evolution of the radial component of the azimuthally
averaged field at the visible solar surface, $B_{r,{R_\odot}}$, we
consider the axisymmetric component of the SFT model \citep[see,
e.g.][]{Cameron:Schuessler:2007,Jiang:etal:2014} given by
\begin{eqnarray}
  \frac{\partial B_{r,{R_\odot}}}{\partial t}&=&
  -\frac{1}{R_\odot \sin\theta}
  \frac{\partial}{\partial \theta}\left(U_{\theta,{R_\odot}}
       B_{r,{R_\odot}} \sin\theta\right) \nonumber \\ 
   & &+\;\frac{\eta_{R_\odot}}{R^2_{\odot} \sin\theta}
     \frac{\partial}{\partial \theta}\left(\sin\theta 
     \frac{\partial B_{r,{R_\odot}}}{\partial \theta}\right)
  + S(\theta,t)\,.
\label{eq:SFT}
\end{eqnarray}
Here $U_{\theta,R_\odot}$ is the poleward meridional flow velocity at
the surface, for which we take $U_{\theta,R_\odot}=-U_0\sin(2\theta)$
with 
$U_0=15\,$m$\cdot$s$^{-1}$ according to observations
\citep[e.g.][]{2010ARA&A..48..289G}.
The surface diffusivity,
$\eta_{R_\odot}$, describes the random walk of magnetic flux elements
transported by supergranular flows \citep{Leighton:1964}.  The source term $S(\theta,t)$
represents the emergence of new flux at the solar surface in the form of
tilted bipolar magnetic regions.  It is convenient to introduce the
quantity
\begin{equation}
a(\theta,t)=\frac{1}{\sin\theta}\int_{0}^{\theta} \sin\theta \,
R_\odot^2 B_{r,R_\odot} \mathrm{d}\theta\,,
\label{eq:def_a}
\end{equation}
which is proportional to the $\phi$-component of the vector potential
for $B_{r,{R_\odot}}$. In terms of $a(\theta,t)$, Eq.~(\ref{eq:SFT}) is
\begin{eqnarray}
  \frac{\partial a}{\partial t}&=&
  -\frac{U_0\sin(2\theta)}{R_\odot\sin\theta} 
   \frac{\partial (a \sin\theta)}{\partial \theta}\nonumber \\
  & & +\frac{\eta_{R_\odot}}{R_\odot^2} \frac{\partial}{\partial \theta}
  \left(\frac{1}{\sin\theta}\frac{\partial
  (a\sin\theta)}{\partial\theta} 
  \right) + a_{\rm S}(\theta,t)\,,
\label{eqn:a}
\end{eqnarray}
where $a_{\rm S}$ is the source term due to flux emergence written in terms of
$a$. Here we take
\begin{equation}
a_{\rm S}(\theta,t) = \alpha \cos\theta \sin^n\theta \, b(\theta,t)\,,
\label{eq:alpha}
\end{equation}
where the proportionality constant $\alpha$ and the integer $n$ are free
parameters of the model. The source term, which is formally similar to
the $\alpha$-effect term in mean-field turbulent dynamo theory, is
assumed to be proportional to the amount of radially integrated toroidal
flux. The factor $\cos\theta$ reflects the latitude-dependence of the
Coriolis force thought to be responsible for the tilt of bipolar
magnetic regions. The factor $\sin^n\theta$ reflects the latitude
dependence of the probability of flux emergence. In most cases we assume
$n=1$, which corresponds to a constant emergence probability per unit
length of the toroidal field lines.

Introducing the definition of $a(\theta,t)$ into Eq.~(\ref{eq:btor_1}),
we obtain the final equation for the integrated toroidal flux, viz.
\begin{eqnarray}
  \label{eqn:b}
  \frac{\partial b}{\partial t}
  &=& \frac{\partial a \sin\theta}{\partial \theta} \epsilon
      \left( \Omega_{R_\odot}- \Omega_{R_{NSSL}} \right)
  -\left(\frac{\partial \Omega_{R_{NSSL}}}{\partial \theta}\right) 
  a \sin\theta \nonumber \\
 & & - \frac{1}{\rsun} \frac{\partial(V_0\sin(2\theta)b)}{\partial\theta}
  + \frac{\eta_0}{R_\odot^2}  \frac{\partial}{\partial \theta} 
 \left[ \frac{1}{\sin \theta}
  \frac{\partial}{\partial \theta} \left(\sin\left(\theta\right) b
  \right) \right]\,
\end{eqnarray}
where the parameter $0 \le \epsilon \le 1$ accounts for the effect of
radial differential rotation below the NSSL. 

Equations~(\ref{eqn:a}) and (\ref{eqn:b}) are the coupled dynamo
equations of our model. Owing to the simplicity of this system, for the
numerical treatment it suffices to use a straightforward
finite-difference scheme with equal spacing in $\theta$.

Equation~(\ref{eqn:a}) contains three parameters. The amplitude of the
poleward meridional flow, $U_0=15\,$m$\cdot$s$^{-1}$ \citep{2010ARA&A..48..289G},
and the turbulent diffusivity at the
surface, $\eta_{\rsun}=250\,$km$^2\cdot$s$^{-1}$ \citep{Komm:etal:1995},
are constrained by observations and SFT
simulations \citep[e.g.,][]{Jiang:etal:2014}. The quantity $\alpha$,
which determines the strength of the dynamo excitation, is difficult to
evaluate empirically and therefore represents a free parameter of the
model.

Eq.~(\ref{eqn:b}) depends on the radial (in the NSSL) and latitudinal
differential rotation given by Eqs.~(\ref{eq:omega_rsun}) and
(\ref{eq:omega_rNSSL}), the effective weighted amplitude of the
equatorward return flow of the meridional circulation, $V_0$, and the
effective turbulent diffusivity, $\eta_0$, for the toroidal field.  The
latter two quantities are treated as free parameters, although some
observational constraints for them exist: a typical speed of about
$1\,$m$\cdot$s$^{-1}$ for the equatorward return flow in low latitudes
may be estimated from the mean latitudinal drift rate of the activity
belts of about $2^\circ$ per year, while a turbulent diffusivity
affecting the toroidal field of
$\simeq$$150$--$450\,$km$^2\cdot$s$^{-1}$ has been obtained by
\citet{Cameron:Schuessler:2016} from properties of the butterfly
diagram. Note that the equatorward transp[ort of the toroidal flux must
not neccesarily be accomplished by a systematic meridional flow:
equatorward convective pumping could provide a similar effect
\citep{Ossendrijver:etal:2002, Hazra:Nandy:2016}.

%$a$$b$$$

In addition, we have the free parameter $0\le\epsilon_{\rm NSSL}\le 1$
which multiplies the radial change of angular velocity in the NSSL, thus
permitting deviations from the assumption that there is no radial
rotational shear below the base of the NSSL. In
Sec.~\ref{sec:parameter_study} we constrain the values of these
parameters (i) by requiring that the dynamo is excited and (ii) by
comparing the results of the model with key observational quantities,
namely the dynamo period and the phase relation between the emergence
rate of bipolar regions (sunspot activity) and the polar field.

\section{Comparison with 2D FTD simulations}
\label{sec:FTD}

In this section we test our model by comparison with results reported in \citet{Karak:Cameron:2016} obtained with the the 2D axisymmetric flux transport dynamo code SURYA, and with a case obtained using the code developed at the MPS and previously used in \citet{2012A&A...542A.127C}  and  \citet{Jiang:etal:2013}.
These studies employ (different) 2D models of the meridional
flow and the differential rotation. The studies of
\citet{Karak:Cameron:2016} uses a simplified analytical form for the
differential rotation mimicking results from helioseismology (see their
Eq.~7) and a one-cell meridional flow profile similar to that of
\citet{Hazra:etal:2014}.  For the 2D run with the code of
\citet{Jiang:etal:2013} we used the meridional flow pattern given in
\citet{Jouve:etal:2008} and a differential rotation according to
\citet{Hathaway:Rightmire:2011} above $r/\rsun=0.94$ and according to
\citet{Schou:etal:1998} below.

The simulations with both 2D codes
include downward convective pumping between the surface and
$r_0=0.88\rsun$ with a constant effective radial velocity $\gamma$, so that
toroidal flux is pushed into the region below $r_0$. The codes use
radial profiles of the magnetic diffusivity that reach an essentially
constant value, $\eta_{\rm CZ}$, below $r_0$.

The relevant value of $\eta_0$ for our model was determined by requiring
that the radially integrated diffusion term (the sixth term in
Eq.~\ref{eq:integrated}) agrees in the 2D case and in our
model. Assuming that $B_\phi$ below $r_0$ does not significantly vary
with radius \citep[which is justified by the results of][see their
Fig.~6]{Karak:Cameron:2016}, this leads to
\begin{equation} 
\eta_0 = \eta_{\rm CZ}\rsun^2 
          {\int_{0.7\rsun}^{0.88\rsun}r^{-1}dr} \left/ 
               {\int_{0.7\rsun}^{0.88\rsun}r dr}\right.
       \simeq 1.609 \eta_{\rm CZ}
\label{eq:eta0}
\end{equation}
The amplitude of the meridional flow at the surface, $U_0$, was
determined by averaging the 2D flow profiles, $v_{\rm 2D}(r,\theta)$,
between $r=\rsun$ and the depth, $d$, for which the Reynolds number of
the pumping, $R_{\rm m}=\gamma d/\eta$, becomes unity, and then taking
the maximum in latitude.  Similarly, the amplitude of the return flow,
$V_0$, was calculated by averaging $v_{\rm 2D}\rsun/r$ for $0.7\le
r/\rsun \le 0.88$ (i.e., up to the bottom of the pumping region) and
taking the maximum in latitude. The latitudinal profiles of both
averaged quantities are reasonably well approximated by $\sin2\theta$.

We consider three cases for the comparison. Cases KC1 and KC2 correspond
to the results shown in Figs.~9 and 11, respectively, of
\citet{Karak:Cameron:2016}. Run KC1 represents a case with high
diffusivity in the convection zone $(100\,$km$^2\cdot$s$^{-1})$, which
reproduces many features of the solar cycle if the flux emergence is
strongly restrictued to low latitudes ($n=12$ in
Eq.~\ref{eq:alpha}). Run KC2 has a two times lower diffusivity, allows
for a broader latitude range of flux emergence ($n=2$), and no return
flow of the meridional circulation. In contrast to case KC1, the
equatorward drift of the toroidal field in this case in due to the
latitudinal propagation of a dynamo wave driven by the radial shear in
the NSSL. Run J carried out with the code used by
\citet{Jiang:etal:2013} is similar to case KC1, albeit with a lower
diffusivity, stronger return flow, and a no artificial restriction of
the latitude range for flux emergence ($n=1$). All 2D runs are linear
with values of the near-surface $\alpha$-effect chosen such that the
dynamos are marginally excited (i.e, growth rate near zero).  Since the
parameter values corresponding to marginal dynamo excitation are not
identical, we performed the calculations with the updated L69 model for
two values of the diffusivity affecting the toroidal field: a) for the
diffusivity $\eta_0$ determined on the basis of $\eta_{\rm CZ}$ as given
by Eq.~\ref{eq:eta0} and b) for an enhanced value, $\eta^*_0$, chosen
such that the dynamo excitation is near marginal.  The parameters used
for the 2D runs and the comparison runs with our updated L69 model are
given in Table~\ref{tab:params}.

\begin{table}[hbt]
  \caption{Parameters for the test cases.}
  \label{tab:params}
\begin{center}
\begin{tabular}{c  c  c  c} \hline
Parameter& KC1& KC2 & J \\ \hline
$\alpha$ [m$\cdot$s$^{-1}$]          & 1.4   & 2.5   & 1.   \\ 
n                                    & 12    & 2     & 1    \\ 
$\eta_{\rsun}$ [km$^2\cdot$s$^{-1}$] & 300.  & 300.  & 250. \\ 
$\eta_{\rm CZ}$[km$^2\cdot$s$^{-1}$] & 100.  & 50.   & 40.  \\ 
$\eta_0$ [km$^2\cdot$s$^{-1}$]      & 160.9 & 80.5  & 64.   \\ 
$\eta^*_0$ [km$^2\cdot$s$^{-1}$]     & 210.  & 210.  & 81.  \\ 
$U_0$ [m$\cdot$s$^{-1}$]              & 15.9  & 15.9  & 15.  \\ 
$V_0$ [m$\cdot$s$^{-1}$]              & 1.6   & 0.    & 4.7  \\ 
$\gamma$ [m$\cdot$s$^{-1}$]          & 35.   & 35.   & 30.  \\ 
%$r_0/R_\odot$                       & .88   & .88    & .88  \\ 
\hline
\end{tabular}
\end{center}
\end{table}

\begin{figure*}
  \centering
  \includegraphics[width=1.\textwidth]{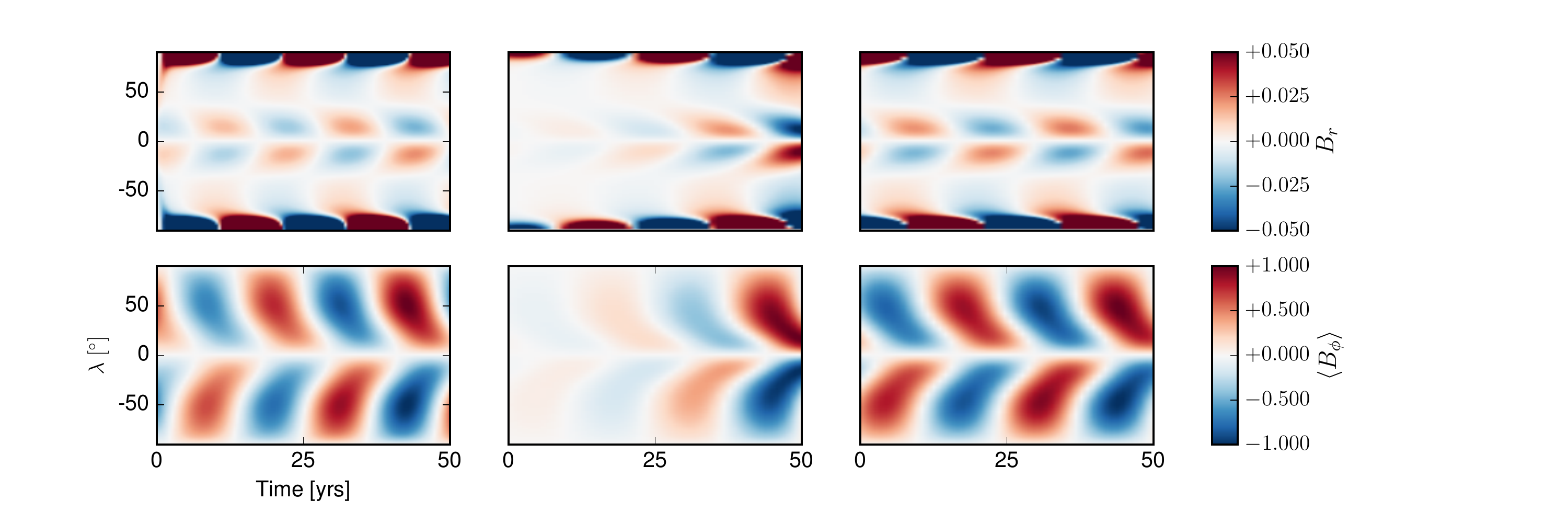}
  \caption{Test case KC1: comparison of the updated Leighton model with
    the results of the 2D dynamo run presented in Fig.~9 of
    \cite{Karak:Cameron:2016}.  The left column gives the results of
    the 2D run, showing time-latitude diagrams for the radial field at
    the surface (upper panel) and the radially averaged toroidal field
    (lower panel). The other two columns show the results of the updated
    Leighton model: using parameters corresponding to those of the 2D model
    (middle column) and with the diffusivity in the convection zone
    increased so that the dynamo has zero linear growth rate (right column).
    The quantities are normalized by a common factor, such that the
    extrema of $\langle B_\phi \rangle$ become $\pm 1$.} 
  \label{fig:KC1}
\end{figure*}

\begin{figure*}
 \centering
  \includegraphics[width=1.\textwidth]{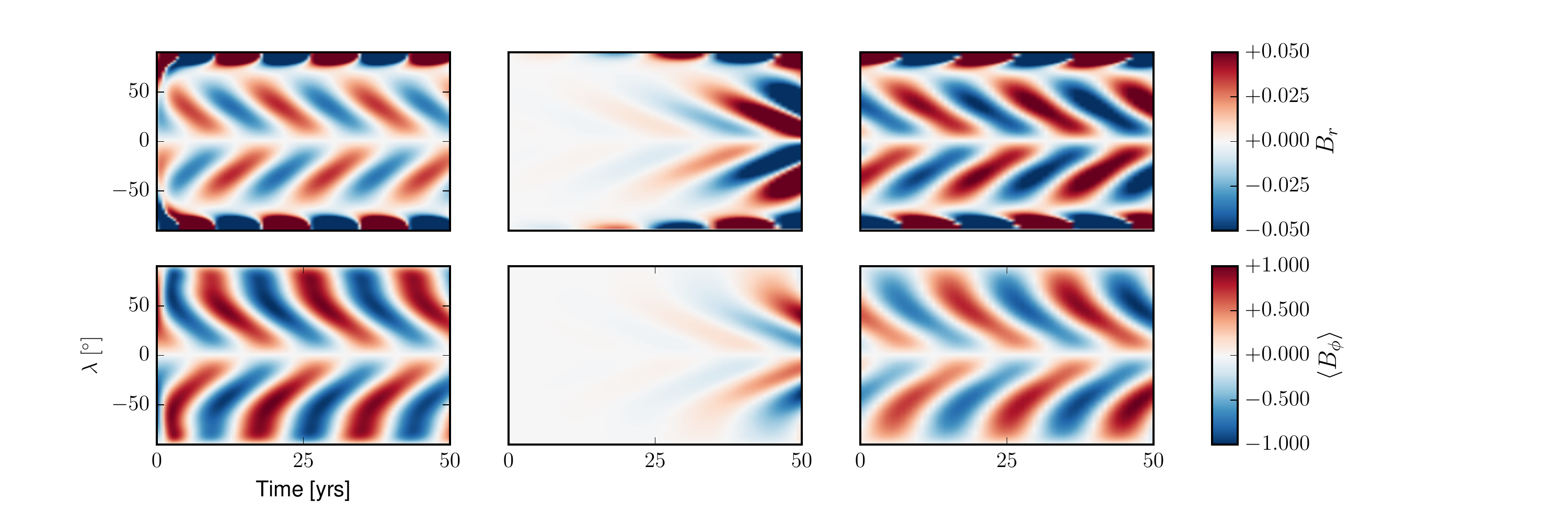}
  \caption{Test case KC2: same as Figure~\ref{fig:KC1}, but for the 2D
  dynamo run shown in Fig.~11 of \cite{Karak:Cameron:2016}.}
  \label{fig:KC2}
\end{figure*}

\begin{figure*}
 \centering
  \includegraphics[width=1.\textwidth]{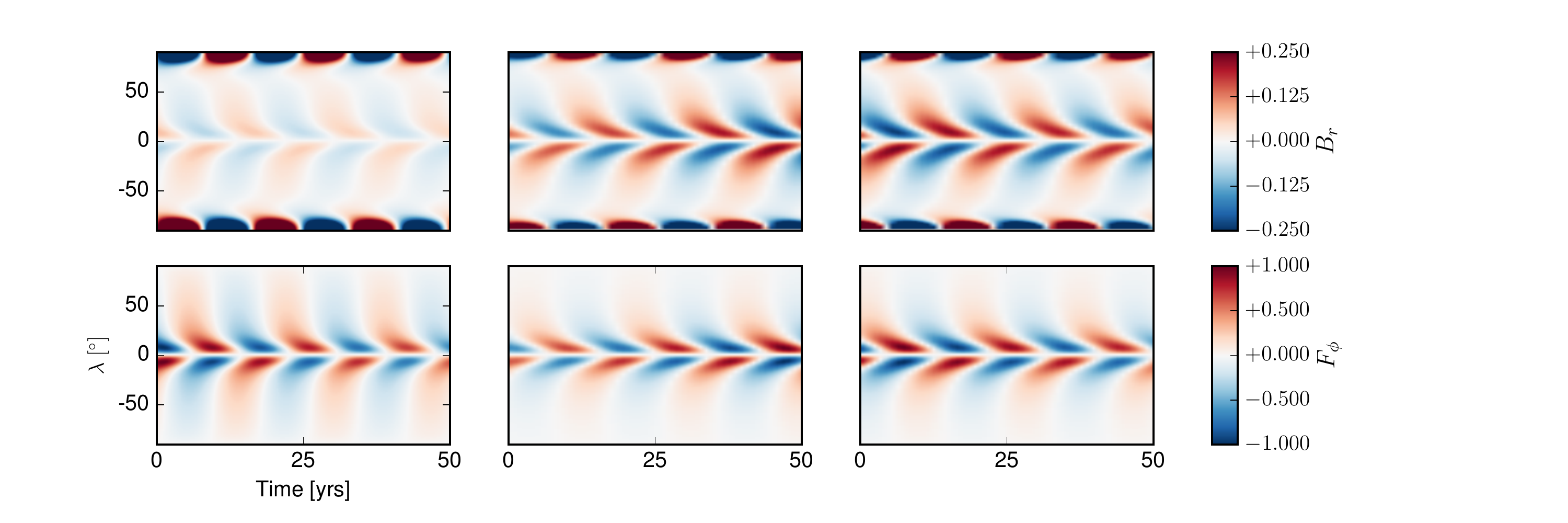}
  \caption{Test case J: same as Figure~\ref{fig:KC1}, but for a
     2D dynamo run carried out with the code used by
    \citet{Jiang:etal:2013}. In this case, the lower row gives the 
    (normalized) radially integrated toroidal flux.}
  \label{fig:J}
\end{figure*}

Figures~\ref{fig:KC1} and \ref{fig:KC2} show the results of the runs
for the cases KC1 and KC2, respectively, while Fig.~\ref{fig:J} gives
the corresponding results for test case J. In all cases, time-latitude
diagrams of the radial field at the surface are represented in the upper
row of the figure. The lower row gives time-latitude diagrams of the
radially averaged toroidal field in cases KC1 and KC2, while the
radially integrated toroidal flux is shown in case J. The 2D results are
shown in the left columns of the figures while the results obatined with the
updated L69 model are given in the middle columns (for diffusivity
$\eta_0$) and in the right columns (for diffusivity $\eta^*_0$ providing
about zero growth rate).

In all cases, the results of the 2D runs and those from our updated L69
model are very similar: dynamo periods, the shape of the butterfly
diagrams, and the phase relation between the polar fields and the
low-latitude toroidal field (representing flux emergence and solar
activity) are reasonably well represented by our much simpler L69
model. There are some minor differences, e.g., the periods are slightly
longer and the low-latitude surface fields somewhat stronger relative to
the polar fields in the L69 cases, but a perfect agreement is obviously
not expected. We thus conclude that the updated L69 model captures the
essential features of these moderately diffusive 2D models.

\section{Parameter study}
\label{sec:parameter_study}

The computational efficiency of the updated L69 model allows us
to systematically cover wide ranges of the parameter values relevant for
the dynamo behaviour. Since many of these parameters (e.g., meridional
flow pattern and magnetic diffusivity in the convection zone, amplitude
of the source term) are uncertain, this is a big advantage in comparison
to the numerically more demanding 2D models. We have thus carried out a
parameter study in order to identify those regions of the parameter
space providing dynamo models that are consistent with basic properties
of the solar cycle. For the latter, we require a) preference of the
(antisymmetric) dipole mode, b) equatorward propagation of the activity
belts, c) cycle period $P\simeq 22\,$years, and d) phase difference
between the maxima of flux emergence (solar activity) and polar field
$\Delta\phi \simeq 90^{\circ}$, meaning that the polar fields
reverse around activity maxima and reach their peak levels around
activity minima.  Of course, the dynamo also must be excited (i.e.,
growth rate $\gamma > 0$).

\begin{figure*}
  \centering
  \includegraphics[width=0.45\textwidth]{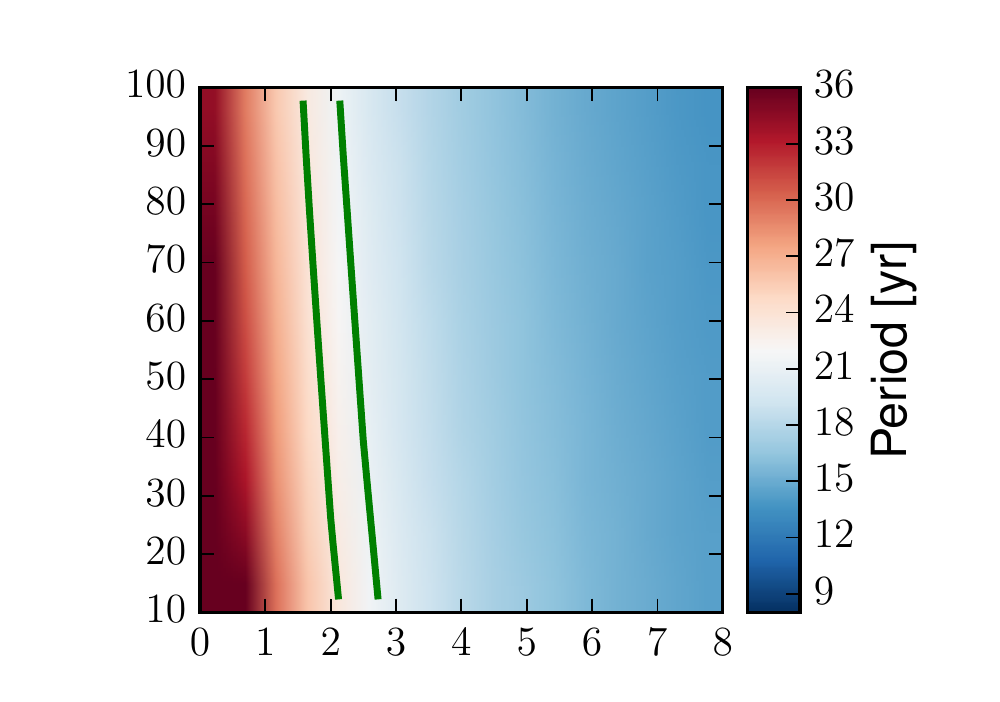}\includegraphics[width=0.45\textwidth]{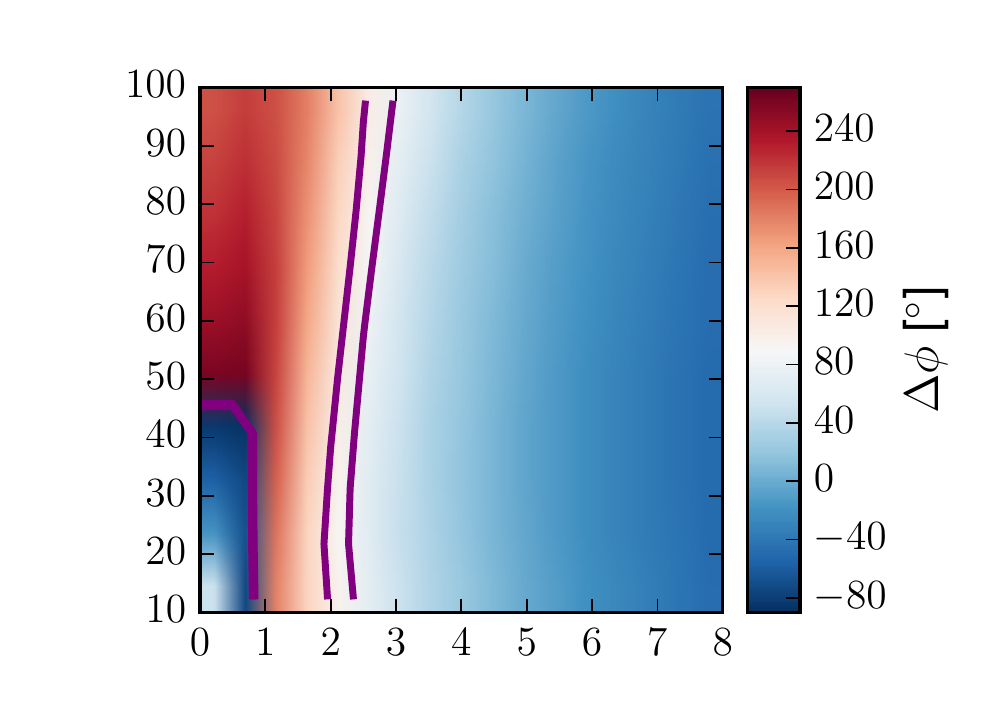}
  \includegraphics[width=0.45\textwidth]{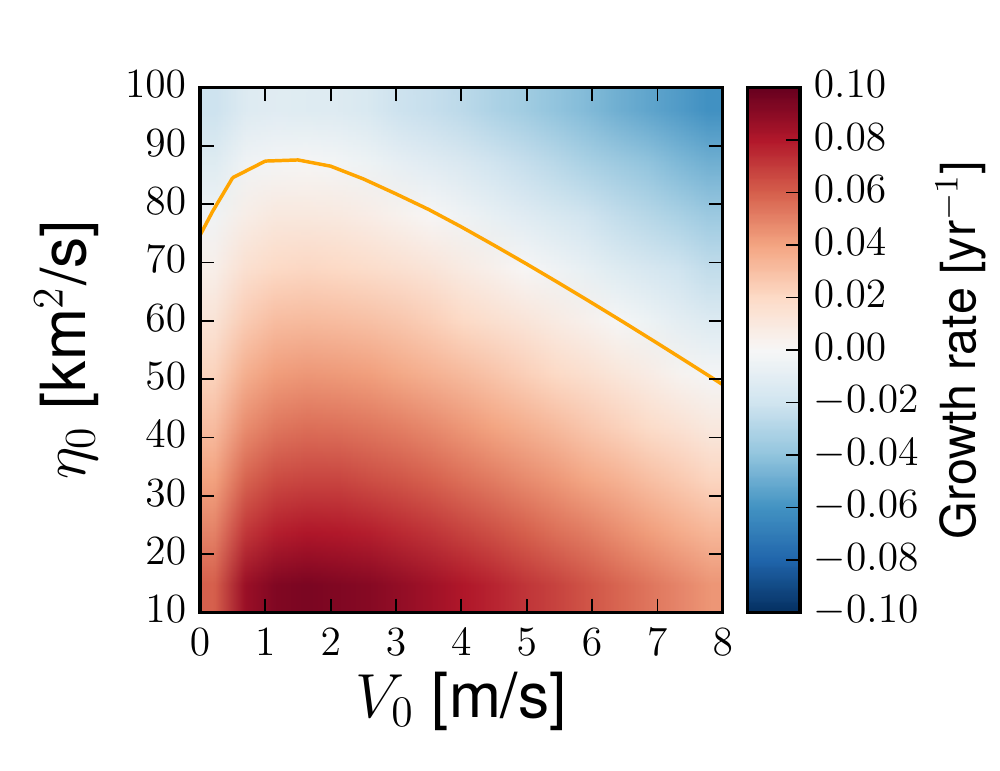}\includegraphics[width=0.45\textwidth]{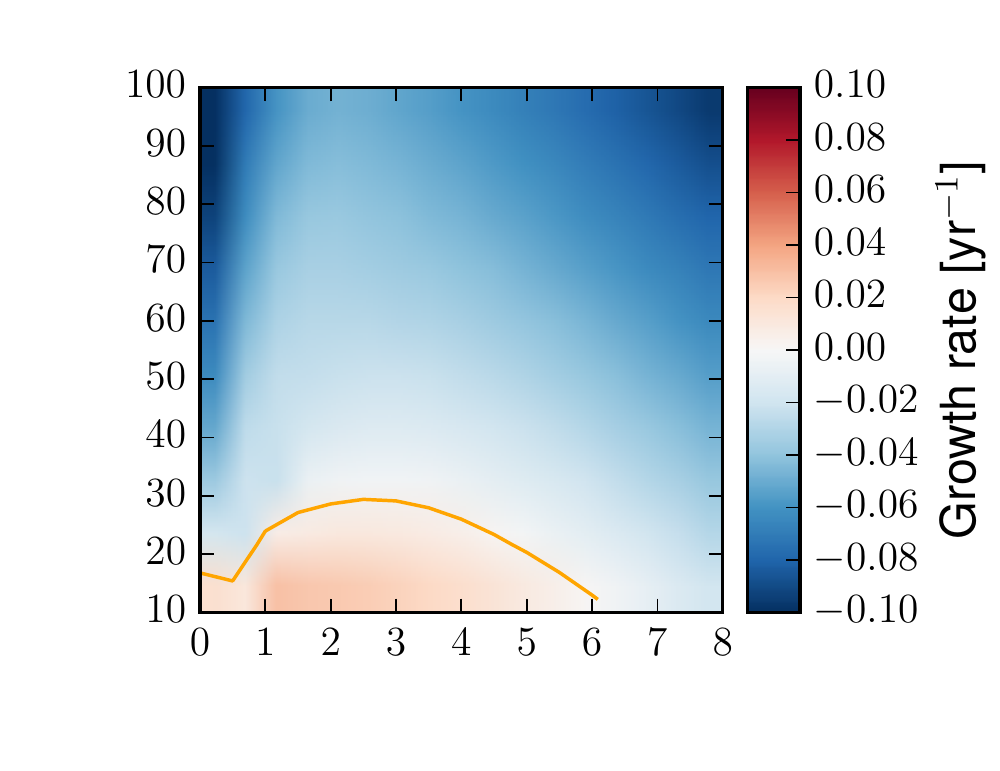}
  \caption{Properties of linear dynamo solutions as functions of the
    amplitude of the effective return meridional flow ($V_0$) and of the
    magnetic diffusivity in the convection zone ($\eta_0$) for values of
    $\alpha=1.4\,$m$\cdot$s$^{-1}$ and $\epsilon=1$. Color shading
    represents the dynamo period ($P$, upper left panel), the phase
    difference between the maximum of the polar field and the maximum
    rate of flux emergence ($\Delta\phi$, upper right panel), and the
    dynamo growth rate ($\gamma$, lower left panel), all for dipole
    parity. The growth rate for quadrupolar parity is given in lower
    right panel. 
    The lines in the lower panels indicate $\gamma=0$,
    thus dividing regions of excited (reddish) and decaying (blueish)
    dynamo solutions.  The lines in the upper panels indicate ranges
    relevant for the solar dynamo: $21\,{\rm yr} \le P \le 23\,{\rm yr}$
    for the period and $80^{\circ}\le \Delta\phi \le 100 ^{\circ}$ for
    the phase difference.}
  \label{fig:parameter2d}
\end{figure*}

A first representation of the results is given in
Fig.~\ref{fig:parameter2d}.  For fixed values of
$\alpha=1.4\,$m$\cdot$s$^{-1}$ and $\epsilon=1.$, the upper panels give
period, $P$, and phase difference, $\Delta\phi$, for dipole parity.  The
lower panels show the dynamo growth rate, $\gamma$, for the dipole
(left) and for the quadrupole (right) modes, respectively.  The lines in
the upper panels indicate ranges relevant for the solar case. The lines
in the lower panels separates regions of excited (below) and non-excited
(above) dynamo solutions.  The figure shows that the updated L69 model
has a clear preference for dipolar parity in the sense that it is
excited in a broader range of diffusivities (i.e., for lower dynamo
number) and that its growth rate is always higher than that of the
corresponding quadrupole mode. Furthermore, the results are consistent
with the period and phase difference in the case of the Sun for
diffusivities below about $80\,{\rm km}^2\cdot$s$^{-1}$ and an effective
return flow of the meridional circulation affecting the toroidal field
in the range 2--3$\,$m$\cdot$s$^{-1}$.

\begin{figure}
  \centering
  \includegraphics[width=0.4\textwidth]{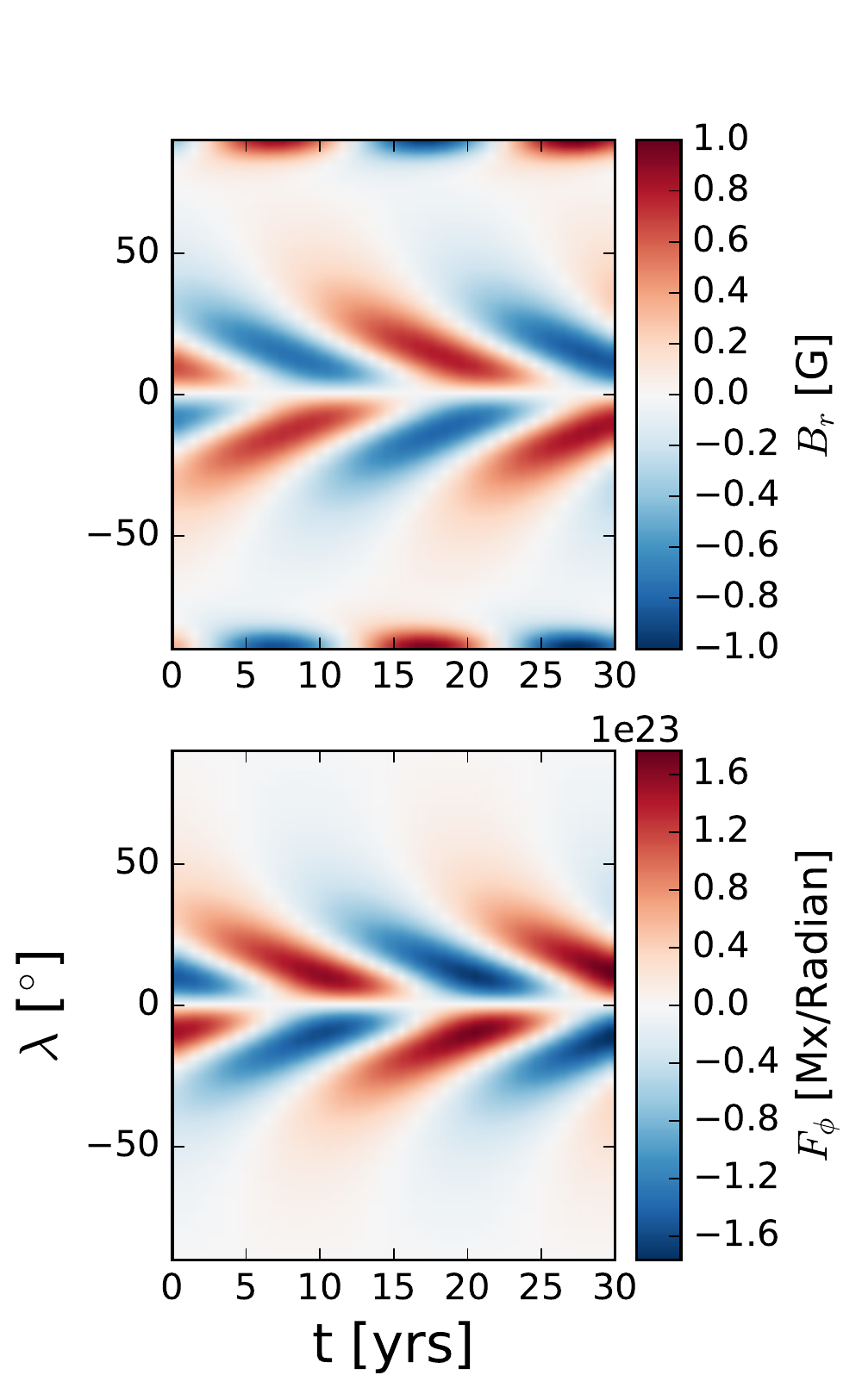}
 \caption{Example of a solar-like dynamo solution for the parameter
 values $\eta_0=80\,$km$^2\cdot$s$^{-1}$,
 $\alpha=1.4\,$m$\cdot$s$^{-1}$, $\epsilon=1$,
 $V_0=2.5\,$m$\cdot$s$^{-1}$. Shown are time-latitude diagrams of the
 radial field at the surface (normalized to a maximum value of 1~G,
 upper panel) and of the radially integrated toroidal magnetic flux 
 (per radian, lower panel).}
 \label{fig:example}
\end{figure}

An example for a solution with solar-like properties is shown in
Fig.~\ref{fig:example}. It has a period of 20.3~years and the correct
phase difference between polar field and flux emergence
$(\Delta\phi=92^\circ)$. Since the amplitude of this linear dynamo model
is arbitrary, we fixed the maxima of the surface field, $B_r$, to
1~G. The corresponding values of the integrated toroidal flux, $F_\phi$,
are consistent with the expected amount of flux residing in the solar
convection zone.  This flux is mainly located in latitudes below
45$^\circ$, so that, in agreement with observation, flux emergence is
also largely restricted to low latitudes. This behaviour results mainly
from the fact that $B_\phi$ is generated by the latitudinal rotational
shear, which is maximal in mid latitudes, together with the equatorward
flux transport by the meridional return flow. Flux-transport dynamos
relying on radial differential rotation near the bottom of the
convection zone, on the other hand, face the problem that the generation
of the toroidal flux mainly takes place in high latitudes
\citep[e.g.][]{Nandy:Choudhuri:2002}.

\begin{figure*}
  \centering
  \includegraphics[width=0.9\textwidth]{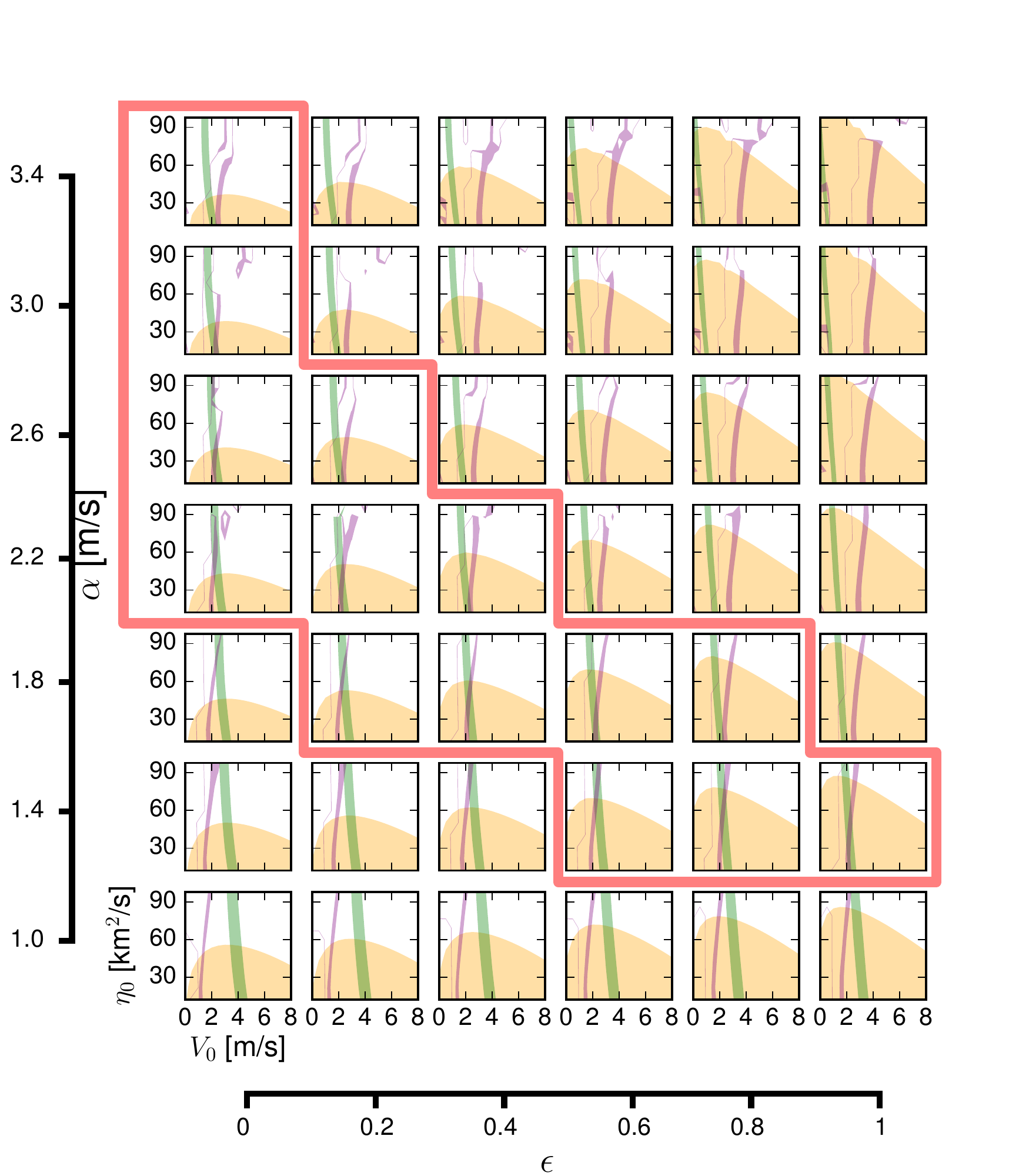}
  \caption{Results of the complete parameter study, for which all four
    parameters describing the subsurface dynamics were varied.  For each
    set of parameters, a simulation was performed and the dynamo period,
    growth rate, and phase difference between the emergence rate and
    polar fields was determined. The green bands indicate the range of
    periods between 21 and 23 years, the purple bands the range of the
    phase difference between $80^{\circ}$ and $100^{\circ}$.  The
    regions shaded in orange indicate dynamo excitation (positive growth
    rate). The red outline encloses those panels where solutions
    matching all three constrains exist.}
  \label{fig:parameter4d_lin}
\end{figure*}

The full results of our parameter study are summarized in
Fig.~\ref{fig:parameter4d_lin}. Here we show only the results for dipole
parity because it invariably is the preferred mode.  The four parameters
describing the subsurface dynamics that we consider are the source
amplitude, $\alpha$, the diffusivity, $\eta_0$, the effective return
flow amplitude, $V_0$, and the quantity $0 \le \epsilon \le 1$, which
represents radial differential rotation below the near-surface shear
layer. We represent the results in a similar form as in
Fig.~\ref{fig:parameter2d} with panels indicating phase difference,
period, and dynamo excitation as functions of $V_0$ and $\eta_0$. We now
give a number of such panels for various values of $\alpha$ and
$\epsilon$, which are indicated by the black bars at the bottom and at
the left side of the figure. The individual panels represent the dynamo
growth rate (positive in the orange shaded regions) and indicate 
relevant `solar' ranges for the dynamo period ($P$ between 21 und 23
years, green bands) and the phase difference ($\Delta\phi$ between
$80^\circ$ and $100^\circ$, purple bands). For each panel, we have
calculated 153 models (using 17 values for $V_0$ in the range
0--8$\,$m$\cdot$s$^{-1}$ and 9 values for $\eta_0$ in the range
10--90$\,$km$^2\cdot$s$^{-1}$).

Fig.~\ref{fig:parameter4d_lin} shows that the domain of excited dynamos
(positive growth rate) grows for increasing values of $\epsilon$,
reflecting the contribution of the NSSL to dynamo excitation. The band
of `solar' values for the phase difference between maximum polar field
and maximum flux emergence (purple bands) is not strongly affected by
changes in $\alpha$, $\eta_0$, and $\epsilon$: in the domain of excited
dynamos, permitted values are in most cases reached for return flow
amplitudes in the range 2--4$\,$m$\cdot$s$^{-1}$. Similarly, the dynamo
period lies in the solar regime for the same range of return flow
velocities. For the highest values of $\alpha$ and not too low
$\epsilon$, periods around 22 years are only reached for very slow
return flow, indicating that the character of the dynamo in this regime
(represented by the panels located in the upper right part of the
figure) has changed from transport-dominated to a dynamo wave driven by
the NSSL. However, in these cases, the phase difference does not match
the required value around 90$^\circ$, so that they cannot be considered
as possible models for the solar dynamo.

The panels enclosed by the thick red contour in
Fig.~\ref{fig:parameter4d_lin} show an overlap of the bands of `solar'
values for period and phase difference. For lower values of $\epsilon$
(less effect of the NSSL), higher values of $\alpha$ are required to
achieve solar-like behaviour.  In the overlap regions, the return flow
amplitude typically has a speed of 2--3$\,$m$\cdot$s$^{-1}$. Considering
the sinusoidal profile of the return flow $(\propto\sin 2\theta)$ with
the extrema reached at $\pm 45^\circ$ latitude, this value is consistent
with the observed latitudinal propagation of the activity belts of about
$2^\circ$ per year (corresponding to about 1$\,$m$\cdot$s$^{-1}$). The
overlap regions are typically located not far away from the line of
marginal dynamo excitation (border of the orange region).
Table~\ref{tab:overlap} gives the parameter ranges for which the
simulations match the solar values of the period, phase difference, and growth rate.

\begin{table}[hbt]
  \caption{Ranges of $\eta$ and $V_0$ for which the different observational constraints
    (on period, phase and growth rate) overlap in Figure~\ref{fig:parameter4d_lin}, for different
    values of $\alpha$, and $\epsilon$. The growth rates, phases, and periods from the grid of simulations
    were interpolated to a resolution in $V_0$ of 0.05~m$\cdot$s$^{-1}$.}
  \begin{center}
    \begin{tabular}{c c|c c} \hline
      $\alpha$ [m$\cdot$s$^{-1}$]& $\epsilon$ & $\eta$ [km$^2\cdot$s$^{-1}$] & $V_0$ [m$\cdot$s$^{-1}$] \\ \hline
      1.4          & 0.6    & 60 --  70   & 2.10-2.40   \\
      1.4          & 0.8    & 40 --  70   & 2.05-2.45   \\
      1.4          & 1.0    & 10 --  70   & 2.00-2.40   \\
      1.8          & 0.2    & 50          & 2.20-2.25   \\
      1.8          & 0.4    & 20 --  60   & 2.05-2.45   \\
      1.8          & 0.6    & 10 --  40   & 2.05-2.40   \\
      1.8          & 0.8    & 10 --  20   & 2.20-2.30   \\
      2.2          & 0.0    & 40          & 2.20-2.25   \\
      2.2          & 0.2    & 10 --  40   & 2.05-2.45   \\
      2.2          & 0.4    & 10 --  50   & 1.95-2.40   \\
      2.6          & 0.0    & 10 --  30   & 2.05-2.35   \\
      2.6          & 0.2    & 10 --  20   & 2.20-2.45   \\
      3.0          & 0.0    & 10 --  30   & 2.15-2.50   \\
      3.4          & 0.0    & 10          & 2.30-2.35   \\
          \hline
    \end{tabular}
  \end{center}
  \label{tab:overlap}
\end{table}

\section{Discussion an conclusions}
\label{sec:conclusions}

We have shown that the dynamo model of \citet{Leighton:1969} can be
updated to include further relevant ingredients (meridional circulation,
convective pumping, near-surface shear layer), so that the results are
consistent with those of more involved 2D flux transport dynamo
models. The uncertainties of the structure of magnetic field and flows
in the convection zone can be condensed into a few free parameters while
the computational simplicity of the model allows us to systematically
scan the associated parameter space. Requiring some essential properties
of the solutions (such as period, parity, phase relation between flux
emergence and polar fields, positive linear growth rate) to agree with
their observed solar counterparts, we were able to strongly narrow down
the parameter space relevant for the solar dynamo. 

We find that the Sun most probably hosts a flux-transport dynamo (as
opposed to a dynamo wave driven by the NSSL) operating not too far from
the threshold of marginal excitation. The latter property is consistent
with recent results for other active stars \citep{Saders:etal:2016,
Metcalfe:etal:2016}. The effective equatorward return flow amplitude for
the toroidal flux (at whichever depth the flux is located in the
convection zone) should be around 2$\,$m$\cdot$s$^{-1}$, which is
consistent with the latitudinal drift rate of the activity belts. Solar
properties are achieved for values of the effective magnetic diffusivity
for the toroidal flux as high as $80\,{\rm km}^2\cdot$s$^{-1}$, which
puts the dynamo in the class of `high-diffusivity dynamos'
\citep[e.g.][]{Choudhuri:2015}. High diffusivity is also indicated by
the observed properties of the solar activity belts
\citep{Cameron:Schuessler:2016}.

The assumption that the tachocline shear is mostly irrelevant for the
generation of toroidal magnetic flux \citep{Spruit:2011,
Cameron:Schuessler:2015, 2016Natur.535..526W} leads to toroidal field and flux emergence
concentrated in low latitudes (as observed) in a natural way: the
rotational shear due to latitudinal differential rotation peaks at mid
latitudes and the deep meridional return flow transports the toroidal
flux equatorwards.

As developed here, our updated L69 model is still linear. In reality,
the cycle amplitudes are limited by a nonlinear back-reaction of the
magnetic field on its sources. There are various possibilities for such
a nonlinearity. While significant suppression of the rotational shear
seems unlikely given the small variations of solar rotation during the
activity cycle \citep{Howe:2009}, the most promising candidate appears
to be a back-reaction on the active-region tilt angle, which is central
ingredient of the poloidal field source. \citet{Dasi:etal:2010} and
\citet{McClintock:Norton:2013} indeed found indications for such an
effect by analysing historical sunspot data \citep[see,
however][]{Wang:etal:2015}. Possibilities for the physical mechanism are
enhanced resistance of stronger fields against the Coriolis force
(before emergence), thermal effects near the base of the convection zone
\citep{Isik:2015}, or the effect of active-region horizontal inflows
\citep{Cameron:Schuessler:2012, Martin-Belda:Cameron:2016}. Nonlinear
effects could modify the parameter space identified here for the
operation of the solar dynamo. However, given that the dynamo probably
operates near the excitation threshold, we do not expect very strong
nonlinear effects.

In addition to scanning a large parameter space, the simplicity and
computational efficiency of the quasi-1D L69 model also allow us to
perform computations covering thousands of cycles. In a forthcoming
paper, we will exploit this property to study how random variations of
the source term affect the variability of the solar cycle over
long time scales.

\bibliographystyle{aa}
\bibliography{BL}

\begin{acknowledgements}
   
\end{acknowledgements}

\begin{appendix}

\section{A violation of the Parker-Yoshimura rule?}
\label{appendix}
The first case discussed in the results section of \citet[][henceforth
L69]{Leighton:1969} involves a model with purely latitudinal
differential rotation. It shows oscillatory solutions with {\em
latitudinally} propagating dynamo waves, leading to solar-like butterfly
diagrams of the toroidal and radial field components (cf. Fig.~1 and 2
of L69). However, if indeed Leighton's model is mathematically
equivalent to the $\alpha\Omega$ formalism as indicated by
\citet{Stix:1974}, then dynamo waves should always propagate along
isorotation surfaces \citep{Yoshimura:1975}. Since this property can be
already be inferred from the Cartesian model of \citet{Parker:1955}, it
is commonly known as the Parker-Yoshimura rule. Consequently, a purely
latitudinal gradient of the angular velocity should lead to radially
propagating dynamo waves \citep[see, e.g.,][]{Koehler:1973} and no
latitudinal propagation, in striking contrast to Leighton's result.

The origin of this apparent contradiction results from an error in
Eq.~(9) of L69, 
\begin{eqnarray}
\label{eq:B1}
  \frac{\partial B_r}{\partial t}&=& 
    -\delta(B_\phi) \frac{FH}{80 \rsun \tau} \frac{\partial}{\partial\mu}
    \left(\mu B_\phi\right)\nonumber \\ 
  & &+\frac{1}{T_D} \frac{\partial}{\partial\mu}\left((1-\mu^2) 
    \frac{\partial B_r}{\partial\mu}\right) \,,
\end{eqnarray}
where $\mu=\cos\theta$ and $F$, $H$, $T_D$, and $\tau$ are parameters
of the model. The critical quantity is the function $\delta(B_\phi)$,
which expresses the assumption that bipolar regions contributing to the
regeneration of the poloidal field are only formed if the toroidal field
exceeds a threshold value, $B_c$:
\begin{equation}
  \delta(B_\phi) = \left\{ \begin{array}{ll} 1 & {\rm for}\; 
         B_\phi \geq |B_c| \\ 0 & {\rm else} \end{array} \right. \,.
\label{eq:delta}
\end{equation}  
Since $B_\phi$ depends on $\theta$, this means that $\delta(B_\phi)$ is
an implicit function of $\theta$. Consistent with the double-ring
formalism of L69, this quantity must therefore be placed within the
differentiation operator in the first term on the r.h.s. of
Eq.~(\ref{eq:B1}), so that this term, ${\cal R}$, should correctly read
\begin{equation}
  {\cal R} = \frac{FH}{80 \rsun \tau} \frac{\partial}{\partial\mu}
                \left[\mu\, \delta(B_\phi) B_\phi\right] \,.
\label{eq:reg}
\end{equation}  
Only for constant $\delta$ (i.e., $B_c=0$), as apparently also assumed
by \citet{Stix:1974}, is Leighton's formulation correct (and consistent
with the $\alpha\Omega$ formalism). For $B_c\neq 0$,
the regeneration term as written in Eq.~(\ref{eq:B1}) 
leads to unphysical results. This can be most easily seen by 
considering the quantity 
\begin{equation}
  a(\theta,t)=\frac{1}{\sin\theta}\int_{0}^{\theta} 
     \sin\theta\, R^2 B_{r,{R_\odot}} \mathrm{d}\theta \,,
\label{eq:vectpot}
\end{equation}  
which is proportional to the vector potential for $B_r$. Integrating
Eq.~(\ref{eq:B1}), we obtain
\begin{eqnarray}
  \frac{\partial a}{\partial t}&=& -\frac{FH\rsun/80}{\tau\sin\theta} 
        \left[\delta(B_\phi)B_\phi\cos\theta +
        \int_0^\theta \frac{\partial \delta}{\partial \theta} 
         B_\phi \cos\theta\, \mathrm{d}\theta \right] \nonumber \\
         & &+\frac{1}{T_D} \frac{\partial}{\partial \theta}
         \left(\frac{1}{\sin\theta}\frac{\partial (a\sin\theta)}
         {\partial\theta} \right) \,.
\label{eq:aa}
\end{eqnarray}
In the language of \cite{Yoshimura:1975}, the
first term on the r.h.s. of Eq.~(\ref{eq:aa})
represents the `regeneration action' (see his Eq.~2.1). In this case, the
term contains a spatial derivative (i.e., is not real in Yoshimura's sense),
so that his theorem does not apply and latitudinal migration is not
excluded for purely latitudinal differential rotation.

Moreover, the incorrect term in Eq.~(\ref{eq:B1}) violates
$\nabla\cdot{\bf B}=0$, which requires $\partial(a\sin\theta)/\partial
t=0$ for $\theta=\pi$ at all times, so that the radial flux integrated
over the entire solar surface vanishes. This condition is trivially
satisfied for the first term in the angular brackets on the r.h.s. of
Eq.~(\ref{eq:aa}) since $B_\phi=0$ at the poles, but the second term
involving $\partial\delta/\partial\theta$ does not necessarily vanish
for $\theta=\pi$.  In the case of the Heaviside function used by
Leighton (see Eq.~\ref{eq:delta}) we have
\begin{eqnarray}
  \int_0^{\pi} \frac{\partial \delta}{\partial \theta} 
  B_\phi \cos\theta\, \mathrm{d}\theta=
  B_c\sum_i (\pm \cos\theta_i)
\end{eqnarray}
where the sum is over the points where $|B_\phi|=B_c$ and the sign
depends on whether $\delta$ jumps from 0 to 1 or vice versa. Clearly,
this sum in most cases does not vanish, so that $\nabla \cdot {\bf{B}}$
is not guaranteed.  Numerical experiments show that the dynamo solutions
with no radial shear reported in L69 decay when the correct form of
Eq.~(\ref{eq:B1}), which maintains the divergence condition, is used.

\end{appendix}

\end{document}